\documentclass[10pt, journal, compsoc]{IEEEtran}

\usepackage[nocompress]{cite}
\usepackage{amsmath,amssymb,xfrac}
\usepackage{algorithm,algpseudocode}
\usepackage{booktabs,multirow}

\newcommand*\ass[1]{[\![#1]\!]}
\newcommand*\mss[1]{\langle#1\rangle}

\newtheorem{theorem}{Theorem}
\newtheorem{lemma}{Lemma}
\newtheorem{definition}{Definition}

\begin{document}

\title{Efficient Privacy-Preserving Computation \\ Based on Additive Secret Sharing}

\author{Lizhi~Xiong,
        Wenhao~Zhou,
        Zhihua~Xia,~\IEEEmembership{Member,IEEE},
        Qi~Gu,
        and Jian Weng,~\IEEEmembership{Member,IEEE}
\IEEEcompsocitemizethanks{
    \IEEEcompsocthanksitem L. Xiong (email: xionglz@nuist.edu.cn), W. Zhou (email: apurance@gmail.com), Z. Xia (corresponding author, email: xia\_zhihua@163.com) and Q. Gu are with the School of Computer and Software,~Nanjing University of Information science and Technology,~Nanjing, China.
    \IEEEcompsocthanksitem J. Weng is with Jinan University, Guangzhou 510632, China.}}


\IEEEtitleabstractindextext{
\begin{abstract}
  The emergence of cloud computing provides a new computing paradigm for users---massive and complex computing tasks can  be outsourced to cloud servers. However, the privacy issues also follow. Fully homomorphic encryption shows great potential in privacy-preserving computation, yet it is not ready for practice. At present, secure multiparty computation (MPC) remains mainly approach to deal with sensitive data. In this paper, following the secret sharing based MPC paradigm, we propose a secure 2-party computation scheme, in which cloud servers can securely evaluate functions with high efficiency. We first propose the multiplicative secret sharing (MSS) based on typical additive secret sharing (ASS). Then, we design protocols to switch shared secret between MSS and ASS, based on which a series of protocols for comparison and nearly all of the elementary functions are proposed. We prove that all the proposed protocols are \textit{Universally Composable} secure in the honest-but-curious model. Finally, we will show the remarkable progress of our protocols on both communication efficiency and functionality completeness.
\end{abstract}

}

\maketitle

\IEEEraisesectionheading{\section{Introduction}\label{sec:introduction}}

\IEEEPARstart{C}{loud} computing as a new computing and data processing paradigm has significantly influenced the way we live. By aggregating the virtualized and interconnected computers and other IT infrastructures, cloud computing provides high-quality Internet service with characteristics of on-demand, broad network access, resource pooling, rapid elasticity, and measured service for users \cite{hamdaqa2012cloud, mell2011nist}. More concretely, for individuals, cloud computing reduces the threshold of accessing computing resources, eliminates obstacles brought by infrastructures, and makes possible complex tasks such as deep learning \cite{shokri2015privacy}. For groups and companies, cloud computing gives more flexibility in allocating resources, make them concentrate more on their business, and cuts expenses. In the past decade, the needs of economy constantly promoted the development of cloud computing, which in turn played an important role in integrating social resources and improving social efficiency.

However, behind such a perfect cooperation model, there are some privacy concerns that cannot be ignored \cite{ryan2011cloud}. In the cloud computing scenario, users inevitably have to hand over data which contain sensitive information to the cloud server. This means that cloud servers can access our private data whenever and wherever they like. And as it turns out, the rise of more and more data breach issues, such as the massive data leakage of Facebook impacting 267 million users in 2017, attracts more attention from the public and drives the implementation of relevant laws \cite{Mancuso2019}. For instance, General Data Privacy Regulation (GDPR), the most comprehensive and widely applied privacy regulation, implemented in the European Union in 2018, is geared towards protecting user privacy and against infringement of data breach. Nevertheless, merely by virtue of socio-legal, it is insufficient to provide truly convincing cloud computing security. Therefore, recent researches on privacy-preserving computation offered a cryptography way to achieve this and it was shown on feasible in theoretical.

The most direct idea is to use Homomorphic Encryption (HE) which happens to coincide with the requirement of cloud computing security: allows performing a series of certain functions on ciphertext with reserving the ability to be decrypted. For example, Paillier \cite{paillier1999public} is one of the partially homomorphic encryption (PHE) with additive homomorphism. Considering two plaintexts $m_1$ and $m_2$, cloud servers can compute $E(m_1+m_2)$ with $E(m_1)$ and $E(m_2)$ without knowing the private key under Paillier encryption scheme. Yet PHE like Paillier allows only one type of operation on ciphertext and thus is not competent for some complex cloud computing tasks. Somewhat homomorphic encryption (SHE) improves the number of types of operations comparing to PHE, such as SYY \cite{sander1999non}, BGN \cite{boneh2005evaluating}, IP \cite{ishai2007evaluating}. These schemes introduce noise to achieve security, resulting in an increase of noise on ciphertext with every evaluation. When the accumulation of noise on ciphertext exceeds a threshold, the ciphertext will lose the ability to be decrypted, which against the definition of HE. Therefore, SHE limits the number of operations on ciphertext and the possibility of fully computation offload. In 2009, a pioneering scheme \cite{gentry2009fully} proposed by Gentry introduced bootstrapping technique to eliminate the noise of ciphertext and achieved the first plausible fully homomorphic encryption (FHE) which can evaluate any operation on ciphertext for any times. In a nutshell, bootstrapping means homomorphically decrypt the ciphertext to get a clean ciphertext that contains no noise and is decryptable. However, bootstrapping is costly and takes tens to hundreds of minutes for every ''clean'' \cite{coron2012public, coron2011fully}. Hence, HE remains a long way to go from theory to practice.

Another way is to use secure multiparty computation (MPC) which allows a group of parties who don't trust each other jointly compute a certain function $\mathcal{F}(x_1,x_2,\cdots,x_n)$ without revealing their private input $x_i$. MPC first introduced by Yao in his seminal Two Millionaires Problem \cite{yao1982protocols} has been developed for four decades and a lot of valuable work for it has emerged. In 1986, Yao proposed a well-known secure 2-party computation (2PC) technic called Garbled Circuits (GC) \cite{yao1986generate} where Alice first generates a garbled Boolean circuit $\mathcal{C}$ which represents the function that Alice and Bob want to perform with their private input, then Bob receives the circuit $\mathcal{C}$  and evaluate it by interacting with Alice. Later, Ben-Or \textit{et al.} proposed BGW protocol \cite{ben2019completeness} which allow parties to evaluate arithmetic circuit consisting of addition and multiplication. In the privacy-preserving computation field, the GMW (aka. secret sharing based MPC) paradigm \cite{goldreich2019play} is usually applied, where users can use secret sharing technic \cite{SHAMIR1979SHARE} to share their private data to servers and then servers evaluate a circuit which represents the function users want to perform. Although Boolean circuit or arithmetic circuit that contains only addition and multiplication is Turing-complete in theoretical, constructing a specific circuit for nonlinear functions to improve efficiency is still a meaningful work. Many efforts to design secure protocols for nonlinear functions such as comparison, division, exponent, etc. have been made \cite{nishide2007multiparty, bogdanov2012high, huang2019lightweight}. However, these schemes either use Taylor series to fit or directly construct Boolean circuit, which not only lead to accuracy loss, but also to low efficiency. The purpose of this paper is to design secure protocols that generate arithmetic circuit for nonlinear functions without Taylor series or bit-level operation as with typical addition and multiplication protocols.

\textbf{Contribution.} In this paper, we propose a secure 2-party computation scheme, which can be applied in privacy-preserving computation under the context of cloud computing. The main contributions can be described as follows:

\begin{itemize}
    \item Inspired by additive secret sharing (ASS) in the context of MPC, we first introduce the multiplicative secret sharing (MSS) into secret sharing based MPC. The shares shared through ASS are with additive homomorphism, while shares shared through MSS are with multiplicative homomorphism. It is almost impossible to give share fully homomorphism, yet it is possible to give share proper homomorphism at the right time.
    \item We design two basic 2PC protocols for switching shared secret between MSS and ASS, including \texttt{SecMulRes} that converts shared secret over MSS to over ASS, and \texttt{SecAddRes} that is the inverse of \texttt{SecMulRes}.
    \item Based on the above two basic protocols, we design a series of 2PC protocols for comparison, exponentiation, logarithm, power, division, product and trigonometric functions, where both the input and output are shared secrets, which follows the GMW paradigm. Finally, we nearly achieve all the elementary functions.
    \item All of these protocols contain only arithmetic operation and no bit-level operation or any fitting. This means that our scheme is not limited to specific floating point precision and avoids the loss of accuracy in theoretical. Moreover, all the protocols are with constant round complexities and low communications.
    \item We prove that all the proposed protocols are Universally Composable secure against the honest-but-curious model.
\end{itemize}

\textbf{Oranization.} The organization of this paper is as follows: We briefly introduce the related work of HE and MPC in Section \ref{sec:relatedwork}. The system model and preliminaries are in Section \ref{sec:preliminaries}. We formally describe secure protocols for linear and multiplication in Section \ref{sec:linear-mul} and our proposed protocols can be found in Section \ref{sec:proposed}. The complexity of protocols is summarized in Section \ref{sec:complexity}. Finally, we give conclusions about our work in Section \ref{sec:conclusions}.

\section{Related Work}\label{sec:relatedwork}

In cloud computing, privacy-preserving computation means that cloud servers can evaluate some functions on ''encrypted'' user data and return ''encrypted'' results to users, where the term ''encrypted'' indicates that the privacy of data is preserved. There are mainly two methodologies which could achieve it: homomorphic encryption and secure multiparty computation.

\textbf{Homomorphic Encryption.} Since the term ''homomorphic encryption'' first proposed by Rivest \textit{et al.} \cite{rivest1978data} in 1978, the purist of FHE has never stopped. As the first attempts of HE, PHE can be divided into two categories: additive HE and multiplicative HE. RSA is the first public key cryptography proposed by Rivest \textit{et al.} \cite{rivest1978method} in 1978, later Rivest \textit{et al.} \cite{rivest1978data} showed the multiplicative homomorphism of RSA and presented the term HE in the same year. Elgamal \cite{elgamal1985public} proposed another widely used multiplicative HE scheme which has higher security than RSA. The additive HE schemes represented by Paillier \cite{paillier1999public} mainly include GM \cite{goldwasser1982probabilistic}, Benaloh \cite{benaloh1994dense}, OU \cite{okamoto1998new}, NS \cite{naccache1998new}, Galbraith \cite{galbraith2002elliptic} and KTX \cite{kawachi2007multi}. Since PHE only supports either addition or multiplication, in some privacy-preserving schemes using PHE \cite{aono2017privacy}, users usually have to undertake part of or even most of the computing tasks, and cloud servers fulfill more of an auxiliary role. Before Gentry's seminal work in 2009, the development of SHE is very limited, where BGN proposed by Boneh \textit{et al.} \cite{boneh2005evaluating} is one of the famous SHE schemes, which allow one multiplication and unlimited addition to being performed on ciphertext. In 2009, Gentry \cite{gentry2009fully} proposed the first FHE scheme in his Ph.D. thesis. Gentry first showed a SHE scheme based on ideal lattices, and then introduced bootstrapping technique (also called decrypt homomorphically) to reduce noise of ciphertext. If a SHE scheme can evaluate its decryption algorithm, then this scheme is bootstrappable and can be transformed to FHE scheme through bootstrapping technique. After Gentry's work, many SHE schemes have been proposed \cite{van2010fully, pisa2012somewhat, brakerski2011fully, brakerski2014efficient}, and bootstrapping also be used to realize FHE. Due to the inefficiency of bootstrapping, Brakerski \textit{et al.} \cite{brakerski2014leveled} proposed a leveled-FHE scheme that can evaluate functions with predefined circuit depth without using bootstrapping. However, the scheme is hardly considered as real FHE, it has better performance than other FHE schemes in some scenario though. At present, FHE is gradually applied in practice and has great efficiency improvement potential.

\textbf{Secure Multiparty Computation.} After Yao presented the Two Millionaires Problem \cite{yao1982protocols}, he proposed GC \cite{yao1986generate} which is one of the first MPC protocols. Later, Beaver \textit{et al.} \cite{beaver1990round} proposed BMR protocols which extend GC from 2PC to MPC by separating Alice's task. Since GC is high costly, many optimization techniques have been proposed to improve efficiency and could be divided into three categories: optimization in GC generation \cite{songhori2015tinygarble, huang2012private}, optimization in GC evaluation \cite{naor1999privacy, beaver1990round, kolesnikov2008improved}, optimization in protocol level \cite{demmler2015aby}. In GC-type schemes, the parties are the private data owner, in other words, they know what input data meaning for. Therefore, Goldreich, Micali, and Wigderson (GMW) \cite{goldreich2019play} presented another MPC paradigm in 1987, where every party first shares his private input (into $n$ shares); then $n$ parties run MPC protocols on shares; finally, $n$ parties recover results from output shares. Specifically, GMW using ASS over $\mathbb{Z}_2$ to share each bit of input, where oblivious transfer \cite{kilian1988founding} is used to evaluate AND gate and XOR gate can be evaluated locally. The GMW paradigm expands the source of MPC input, that is, allowing private input outside the parties such as users. Follow this paradigm, Ben-Or \textit{et al.} \cite{ben2019completeness} proposed BGW protocol which allows parties evaluate arithmetic circuit based on Shamir secret sharing \cite{SHAMIR1979SHARE}, where both addition and multiplication can be evaluated locally. Since multiplication will double the order of polynomial and thus the secret cannot be recovered from shares, BGW introduced degree reduction via resharing shares. However, this method is not only inefficient, but also limits the threshold of Shamir secret sharing requiring that $2t<n$. To solve this problem, Beaver creatively introduced Beaver triple \cite{beaver1991efficient} generated in offline phase to significantly improve efficiency of evaluating multiplication. In 2006, Damg{\aa}rd \textit{et al.} \cite{damgaard2006unconditionally} presented bit-decomposition protocol to convert shared secret over arithmetic sharing to over bitwise sharing. This technique allows us to combine the advantages of Boolean circuit and arithmetic circuit. Since then many MPC schemes based on ASS and bit-decomposition technique have been proposed \cite{bogdanov2008sharemind, bogdanov2012high, demmler2015aby}. In addition, there are many works that are dedicated to implementing MPC protocols and even provide secure language to hide details of MPC protocols \cite{malkhi2004fairplay, bogdanov2008sharemind, demmler2015aby, henecka2010tasty}. Recently, with the rapid development of machine learning, many schemes that use MPC technic to support privacy-preserving for the task have been proposed \cite{wagh2018securenn, mohassel2017secureml, huang2019lightweight}. Most of them optimized MPC they used according to the characteristics of machine learning.

\section{System Model and Preliminaries}\label{sec:preliminaries}

In this section, we describe the GWM paradigm which is the basis of this paper, the threat model we use, definition of security, and the secret sharing technic.

\subsection{GMW Paradigm}

GMW \cite{goldreich2019play} introduced secret sharing in MPC to encrypt private input and become a new MPC paradigm called GMW paradigm. In this paper, we follow the GMW paradigm, and which can be divided into three stages:

\begin{enumerate}
    \item \textit{All private inputs are shared by their owners.} On the one hand, secret sharing encrypts private data, on the other hand, it decouples the secret owner and the sharing holder, i.e., the private data owner and the party. Therefore, secret sharing expands input sources, which means that non-parties, such as users, can provide data to parties, while in MPC scenarios, the private data owner is equal to the party. In this paper, the ASS is set to the default secret sharing technic. Since ASS is an $(n, n)$ threshold scheme, sharing private input entails that every party holds a share of every input.
    \item \textit{The parties evaluate the circuit on shares.} In MPC, parties will generate corresponding circuit according to specific tasks before the execution of protocol, while in the context of cloud computing, since the party (i.e., the server) is the outsourcer, the users need to send the circuit to the parties. Indeed, the users can also send the function to be executed and the circuit will be generated by the parties. Regardless of the data source, the function and its details are clear and public to parties. This means that users can only keep data private to the server but not function private. In this paper, we specify that there are two parties $P_1$ and $P_2$ (i.e., two servers) due to the fact that the fewer servers involved in computing, the more secure it is for users.
    \item \textit{The final result is recovered by output shares.} After the evaluating stage, parties will output shares of the result. In cloud computing, these shares will then be sent back to users. The principal feature of GMW paradigm is that input and output, including intermediate results, are in the form of share, which is similar to the ciphertext form in homomorphic encryption.
\end{enumerate}

In GMW paradigm, the main effort (include this paper) is to design efficient protocols to generate circuit for functions. Moreover, since the users are not involved in computation, the number of users and communication between users and parties are not considered in protocol design.

\subsection{Threat and Security Model}

Similar to most secret sharing based MPC schemes \cite{goldreich2019play, bogdanov2012high, bogdanov2008sharemind}, the \emph{honest-but-curious} model (aka. semi-honest model) is used in this paper, where an honest-but-curious adversary could corrupt parties before the execution of protocol, in other words, each party could be seen as honest-but-curious adversary. An honest-but-curious adversary will abide by the protocol honestly but will analyze all the data he has as much as possible. In addition, we assume that two parties will not collude with each other, which means the one adversary cannot corrupt two parties at the same time.

The real-ideal paradigm firstly presented in \cite{goldwasser1984probabilistic} is widely used to prove security of stand-alone MPC protocol. In the ideal world, there is a trusted third party $\mathcal{T}$ who will receive private input $x_i$ sent by every party to compute $\mathcal{F}(x_1,x_2,\cdots,x_n)$ and send results to parties. The trusted third party $\mathcal{T}$ is also called ideal functionality $\mathcal{F}$, in other words, with the help of $\mathcal{T}$, parties securely compute the function $\mathcal{F}$. In the real world, there is no such $\mathcal{T}$, parties have to interact with each other to compute $\mathcal{F}$ according to protocol. Roughly, a protocol $\pi$ is secure or securely realizes ideal functionality $\mathcal{F}$ if the distributions of input and output in the ideal world and in the real world are indistinguishable.

However, the real-ideal paradigm cannot guarantee the composability of secure protocol, which means that a complex protocol composed of secure sub-protocols is still a stand-alone protocol and should be proved from scratch. The Universally Composable (UC) Security framework proposed by Canetti \cite{canetti2001universally} introduced the environment on the basis of the real-ideal paradigm to give secure protocol with the composability. A sub-protocol can be considered as the ideal functionality in proving security, if it is UC-secure. In this paper, we will prove the security of our protocols under UC framework and the following security definition will be used.

\begin{definition}
    A protocol $\pi$ is UC-secure or UC-securely realizes the ideal functionality $\mathcal{F}$ if for any real world adversary $\mathcal{A}$ there exists a probabilistic polynomial-time simulator $\mathcal{S}$ that can simulate an ideal world view such that is indistinguishable from real world view.
\end{definition}

To prove that a protocol is UC-secure, it suffices to show that the incoming view consisting of the message sent by other parties and output is simulatable for any parties (adversary). In addition, the following lemmas will be used.

\begin{lemma}
    The element $x+r$ is uniformly distributed and independent from $x$ for any element $x\in\mathbb{F}$ if the element $r\in\mathbb{F}$ is also uniformly distributed and independent from $x$ \cite{bogdanov2012high}.
    \label{lemma:add}
\end{lemma}

\begin{IEEEproof}
    Readers can refer to the proof in \cite{bogdanov2012high}.
\end{IEEEproof}

\begin{lemma}
    The nonzero element $x\cdot{}r$ is uniformly distributed and independent from $x$ for any element $x\in\mathbb{F}$ if the element $r\in\mathbb{F}$ is also uniformly distributed and independent from $x$.
    \label{lemma:mul}   
\end{lemma}

\begin{IEEEproof}
    If the nonzero element $r\in\mathbb{F}$ is uniformly distributed and independent from all $x$ then so is $x\cdot{}r$, since $f_r(x):=x\cdot{}r$ is a bijective mapping for $\mathbb{F}$.
\end{IEEEproof}

\subsection{Secret Sharing}

Secret sharing introduced by Shamir \cite{SHAMIR1979SHARE} is aimed at addressing a problem in which the secret data divided into $n$ pieces can be restored with over a certain number of pieces, whereas cannot with less of those. Specifically, in \cite{SHAMIR1979SHARE}, Shamir presented a $(k, n)$ threshold scheme such that: 1) the secret data $s$ which is named \textit{secret} will be securely divided into $n$ pieces $s_i$ which are named \textit{shares}; 2) any $k$ shares are able to reveal complete information about the secret; 3) any $k-1$ or less shares reveals no useful information about the secret. One wants to share a secret $s$, he first securely divides secret into $n$ shares $s_i$ via secret sharing, then sends them to $n$ parties respectively. Finally, any $k$ parties can restore the secret $s$ together, whereas any $k-1$ or less parties cannot.

Additive secret sharing (ASS) is defined over a field $\mathbb{F}$. In ASS, a secret $x\in\mathbb{F}$ will be randomly divided into $n$ shares $[x_1,x_2,\cdots,x_n]$ such that $x_1+x_2+\cdots+x_n=x$. Specially, ASS defines an $(n, n)$ threshold scheme, since the restoration of the secret requires all of the shares. The secret sharing used in GMW protocol \cite{goldreich2019play} can be regarded as ASS defined over a finite field $\mathbb{Z}_2$ and is suitable for Boolean circuit. In \cite{bogdanov2008sharemind, bogdanov2012high, wagh2018securenn, mohassel2017secureml, huang2019lightweight}, ASS defined over a finite ring $\mathbb{Z}_{2^{l}}$ or a finite filed $\mathbb{F}$ is used and most of the secure operation is on arithmetic circuit. The defined field for ASS determines the element form of shares and type of circuit on which parties want to perform. In this paper, the ASS defined over real field $\mathbb{R}$ is considered, which is adequate for comprehensive operations on arithmetic circuit.

Inspired by ASS, we introduce the multiplicative secret sharing (MSS) defined over real field $\mathbb{R}$ for the needs of protocol design. Specifically, in MSS, a secret $u\in\mathbb{R}$ will be randomly divided into $n$ shares $[u_1,u_2,\cdots,u_n]$ such that $u_1\times u_2\times\cdots\times u_n=u$. Similar to ASS, MSS defines an $(n, n)$ threshold scheme. Note that it is not appropriate to use MSS defined over some fields such as $\mathbb{Z}_2$ due to the fact that once a share is $0$, the party could know the secret is $0$.

The original intention of secret sharing is to provide access control for sensitive data. In this case, the threshold $t$ should make a tradeoff between security and reliability, where a higher $t$ means more security and less reliability, while a lower $t$ is the opposite. As an $(n, n)$ threshold scheme, ASS and MSS obviously sacrifice reliability and thus is not suitable in this scenario. However, they provide MPC with capabilities of data encryption and data distribution.

For the sake of brevity, we use $\ass{x}$  to denote all of shares $[x_1,x_2,\cdots,x_n]$ generated by the secret $x$ through ASS, and use $\mss{u}$ to denote all of shares $[u_1\times u_2\times\cdots\times u_n]$ generated by the secret $u$ through MSS. More precisely, the notation $\ass{x}$ or $\mss{u}$ mean share $x_i$ or $u_i$ for party $\mathcal{P}_i$, in other words, it represents an ensemble of individuals.

\section{Linear and Multiplication Protocols}\label{sec:linear-mul}

Secure protocols for addition can be implemented directly based on additive homomorphism of ASS, meanwhile, secure multiplication protocol can be constructed efficiently with Beaver triple \cite{beaver1991efficient}. The implementation of these protocols is not the contribution of this paper, but we still devote a section here to formally describe them for completeness.

As illustrated in Algorithm \ref{alg:seclinear}, secure linear protocol which further considers the operation of multiplication by public number takes $n$ shared secrets $\ass{x^1},\ass{x^2},\cdots,\ass{x^n}$ with their public coefficient $a^1, a^2,\cdots, a^n$ and a public bias $b$ as input, and return the shared secret $\ass{\sum_{j=1}^{n}a^j\cdot{}x^j+b}$. For clarity, superscript here is used to distinguish different secrets or public numbers, whereas subscript is used to represent the share of corresponding party. During the execution of secure linear protocol, $\mathcal{P}_1$ and $\mathcal{P}_2$ are only required to perform same linear operation on their own shares asynchronously.

\begin{theorem}
    Algorithm \ref{alg:seclinear} is correct and UC-secure in the honest-but-curios model.
\end{theorem}

\begin{IEEEproof}
    For correctness, there are
    \begin{align*}
        f = & f_1 + f_2 = \sum_{j=1}^{n}a^j \cdot x_1^j + \sum_{j=1}^{n}a^j \cdot x_2^j + b\\
          = & \sum_{j=1}^{n}a^j \cdot (x_1^j+x_2^j) + b = \sum_{j=1}^{n}a^j\cdot{}x^j+b
    \end{align*}
    For security, note that Algorithm \ref{alg:seclinear} contains only local operations and no interaction with each party. Therefore, the protocol is perfectly simulatable and is UC-secure in the honest-but-curios model.
\end{IEEEproof}

\begin{algorithm}
    \caption{Secure Linear Protocol $\ass{f}\gets SecLinear(\ass{x^1},\ass{x^2},\cdots,\ass{x^n},a^1,a^2,\cdots,a^n,b)$}\label{alg:seclinear}
    \begin{algorithmic}[1]
      \Require
        Shared secrets $\ass{x^j}$, public coefficient $a^j$ for all $j\in\{1,2,\cdots,n\}$ and bais $b$.
      \Ensure
        Shared secret $\ass{f}$ such that $f=\sum_{j=1}^{n}a^j\cdot{}x^j+b$.
      \State $\mathcal{P}_1$ computes $f_1\gets\sum_{j=1}^{n}a^j \cdot x_1^j$.
      \State $\mathcal{P}_2$ computes $f_2\gets\sum_{j=1}^{n}a^j \cdot x_2^j+b$.
      \State \textbf{Return} $\ass{f}$.
    \end{algorithmic}
\end{algorithm}

Unlike linear operation, secure protocol for multiplication cannot be constructed by such intuitive way, since $\ass{x\cdot y}=\ass{x}\cdot\ass{y}$ fails. In 1991, Beaver \cite{beaver1991efficient} presented a crucial work to solve this problem, where a multiplicative triple (aka. Beaver triple) is used to assist parties to securely compute product. The idea of beaver triple has its roots in a simple but inconspicuous mind: obfuscate inputs and then use a particular linear combination to correct them. A Beaver triple consists of three numbers $a,b,c$ such that $c=a\cdot b$ where $a$ and $b$ are random and will be generated and shared by ASS in offline phase. It means that each party $\mathcal{P}_i$ will own shares of input $x_i$ and $y_i$ and shares of Beaver triple $a_i$, $b_i$, and $c_i$ before the execution of the protocol. In general, there are two ways to generate and share Beaver triple in offline phase: 1) done by parties through other MPC technic \cite{mohassel2017secureml}; 2) done by trusted third party $\mathcal{T}$ \cite{huang2019lightweight}. Specially, in privacy-preserving computation, users can act as a trusted third party.

Secure multiplication protocol takes two shared secrets $\ass{x}$ and $\ass{y}$ as input, and return the shared secret $\ass{x\cdot y}$. Firstly, two random number $a$ and $b$ used to mask two multipliers $x$ and $y$ respectively via computing differences $\ass{d}\gets\ass{x}-\ass{a},\ass{e}\gets\ass{y}-\ass{b}$. Secondly, parties reveal $d$ and $e$ to make them into public numbers. Since both $a$ and $b$ are random and privately shared, revealing $d$ and $e$ leaks no information about $x$ and $y$. Now multiply two differences $d$ and $e$ to obtain $d \cdot e = x \cdot y - a \cdot y - b \cdot x + a \cdot b$, where $x \cdot y$ is the result that we want, meanwhile $a \cdot y$, $b \cdot x$, and $a \cdot b$ can be eliminated by $e \cdot \ass{a}$, $d \cdot \ass{b}$, and $\ass{c}$. Finally, it conclude that $\ass{x \cdot y} = d \cdot e + d \cdot \ass{b} + e \cdot \ass{a} + \ass{c}$. Algorithm \ref{alg:secmul} shows the details, where the term $e \cdot d$ can be computed by either party, and we specify $\mathcal{P}_2$ here without loss of generality. Moreover, secure protocol for product which has more than 2 multipliers can be implemented by recursively splitting input sequences in half and then computing each pair in parallel.

\begin{theorem}
    Algorithm \ref{alg:secmul} is correct and UC-secure in the honest-but-curios model.
\end{theorem}

\begin{IEEEproof}
    For correctness, there are
    \begin{align*}
        f = & f_1 + f_2 \\
          = & c_1 + d \cdot b_1 + e \cdot a_1 + c_2 + d \cdot b_2 + e \cdot a_2 + e \cdot d \\
          = & c + d \cdot b + e \cdot a  + e \cdot d = x \cdot y
    \end{align*}
    To prove security, we first consider the incoming view and output of $\mathcal{P}_1$ and prove that both are simulatable. The incoming view of $\mathcal{P}_1$ is $(d_2,e_2)$ and the output is $(f_1)$, where $d_2=x_2-a_2$, $e_2=y_2-b_2$, and $f_1=c_1+d\cdot b_1 + e \cdot a_1$. The values $a_2$, $b_2$, and $c_1$ are uniformly distributed and independent from any private input, due to Lemma \ref{lemma:add}, the incoming view and output are simulatable. Since $\mathcal{P}_1$ and $\mathcal{P}_2$ are symmetric in Algorithm \ref{alg:secmul}, it is trivial to build a simulator $\mathcal{S}$ for $\mathcal{P}_2$.
\end{IEEEproof}

\begin{algorithm}[t]
    \caption{Secure Multiplication Protocol $\ass{f}\gets SecMul(\ass{x},\ass{y})$}\label{alg:secmul}
    \begin{algorithmic}[1]
      \Require
        \Statex Shared secrets $\ass{x}$ and $\ass{y}$.
        \Statex Beaver triple: $\ass{a},\ass{b},\ass{c}$, where $c = a \cdot b$. 
      \Ensure
        Shared secret $\ass{f}$ such that $f=x\cdot y$.
      \State $\mathcal{P}_1$ computes $d_1\gets x_1-a_1, e_1\gets y_1-b_1$.
      \State $\mathcal{P}_2$ computes $d_2\gets x_2-a_2, e_2\gets y_2-b_2$.
      \State $\mathcal{P}_1$ and $\mathcal{P}_2$ collaboratively reveal $d$ and $e$.
      \State $\mathcal{P}_1$ computes $f_1\gets c_1 + d \cdot b_1 + e \cdot a_1$.
      \State $\mathcal{P}_2$ computes $f_2\gets c_2 + d \cdot b_2 + e \cdot a_2 + e \cdot d$.
      \State \textbf{Return} $\ass{f}$
    \end{algorithmic}
\end{algorithm}

\section{Proposed Protocols}\label{sec:proposed}

In this section, we firstly propose two resharing protocols. Then, based on these, we design a series of secure protocols for comparison, exponentiation, logarithm, power, and trigonometric functions. Finally, we discuss the remainder of basic elementary functions---inverse trigonometric functions.

\subsection{Resharing}

MSS described in Preliminaries has multiplicative homomorphism different from ASS, and in fact, is the basis of all the remaining protocols. However, all data (secret) should eventually return to the form shared by ASS in our setting, hence the resharing technic which switches shared secret between MSS and ASS need to be first implemented. Besides, based on MSS, we can directly construct secure protocol for comparison.

\subsubsection{Multiplicative Resharing}

The multiplicative resharing is to convert shared secret over MSS to over ASS. In fact, it is homologous with of multiplication. Recall that multiplication is essentially to compute $(f_1+f_2)\gets (x_1+x_2)\cdot(y_1+y_2)$, where the inputs of $\mathcal{P}_i$ are $x_i$ and $y_i$; whereas multiplicative resharing is essentially to compute $(x_1+x_2)\gets(u_1\cdot u_2)$, where the input of $\mathcal{P}_i$ is $u_i$. The outputs of both are isomorphic, while in terms of input, $u_1$ can be considered as $(x_1+x_2)$ and $u_2$ can be considered as $(y_1+y_2)$. That is to say, for $\mathcal{P}_1$, he has $x_1$ and $x_2$ instead of $x_1$ and $y_1$, thereby Beaver triple allocation needs to change accordingly. Specifically, a Beaver triple will be generated and the value $c$ will be shared by ASS in offline phase as usual. The difference is that the values $a$ and $b$ will be directly sent to $\mathcal{P}_1$ and $\mathcal{P}_2$ instead of sharing them, since for $\mathcal{P}_1$, he owns $x_1$ and $x_2$ and hopes to get $a_1$ and $a_2$. In other words, before the execution of this protocol, $\mathcal{P}_1$ owns $u_1$, $a$, and $c_1$, meanwhile $\mathcal{P}_2$ owns $u_2$, $b$, and $c_2$. As illustrated in Algorithm \ref{alg:secmulres}, at first, $\mathcal{P}_1$ computes $d\gets u_1-a$ which can be seen as $\ass{e}\gets\ass{y}-\ass{b}$. Then the values $d$ and $e$ are revealed by two parties. At last, $\mathcal{P}_1$ computes $x_1\gets c_1+e\cdot a$ and $\mathcal{P}_2$ computes $x_2\gets c_2+d\cdot b+e\cdot d$ respectively, which can be seen as $\ass{c}+d\cdot\ass{b}+e\cdot\ass{a}+e\cdot d$. Same as secure multiplication protocol, the term $e\cdot d$ is specified here to be computed by $\mathcal{P}_2$.

\begin{theorem}
    Algorithm \ref{alg:secmulres} is correct and UC-secure in the honest-but-curios model.
\end{theorem}

\begin{IEEEproof}
    For correctness, there are
    \begin{align*}
        x = & x_1 + x_2 = c + d \cdot b + e \cdot a + e \cdot d \\
          = & c + u_1 \cdot b - a \cdot b + u_2 \cdot a - a \cdot b \\
            & + u_1 \cdot u_2 - u_2 \cdot a - u_1 \cdot b + a \cdot b \\
          = & u_1 \cdot u_2 = u
    \end{align*}
    To prove security, since $\mathcal{P}_1$ and $\mathcal{P}_2$ are symmetric in Algorithm \ref{alg:secmulres}, it suffices to show that the view of $\mathcal{P}_1$ is simulatable. The incoming view of $\mathcal{P}_1$ is $(e)$ and the output is $(x_1)$, where $e=u_2-b$ and $x_1=c_1+e\cdot a$. Since the values $b$ and $c_1$ are uniformly distributed and independent of any private input, due to Lemma \ref{lemma:add}, it is easy to perfectly build a simulator for $\mathcal{P}_1$. 
\end{IEEEproof}

\begin{algorithm}[t]
    \caption{Secure Multiplicative Resharing Protocol $\ass{x}\gets SecMulRes(\mss{u})$}\label{alg:secmulres}
    \begin{algorithmic}[1]
      \Require
        \Statex Shared secret $\mss{u}$ over MSS.
        \Statex Beaver triple: $a,b,\ass{c}$, where $c = a \cdot b$. The value $c$ is shared by ASS, meanwhile $a$ is held by $\mathcal{P}_1$ and $b$ is held by $\mathcal{P}_2$.
      \Ensure
        Shared secret $\ass{x}$ over ASS such that $x=u$.
      \State $\mathcal{P}_1$ computes $d\gets u_1-a$.
      \State $\mathcal{P}_2$ computes $e\gets u_2-b$.
      \State $\mathcal{P}_1$ sends $d$ to $\mathcal{P}_2$.
      \State $\mathcal{P}_2$ sends $e$ to $\mathcal{P}_1$.
      \State $\mathcal{P}_1$ computes $x_1\gets c_1 + e \cdot a$.
      \State $\mathcal{P}_2$ computes $x_2\gets c_2 + d \cdot b + e \cdot d$.
      \State \textbf{Return} $\ass{x}$
    \end{algorithmic}
\end{algorithm}

\subsubsection{Additive Resharing}

Additive resharing which converts shared secret over ASS to over MSS is defined as the inverse of multiplicative resharing and secure protocol for it can be constructed by following Algorithm \ref{alg:secmulres} backward step by step. In offline phase, the allocation of Beaver triple is the same as Algorithm \ref{alg:secmulres}. In online phase, $\mathcal{P}_1$ and $\mathcal{P}_2$ first compute $\ass{t}=\ass{x}-\ass{c}$. Note that since the term $e\cdot d$ is assumed to be done by $\ass{P}_2$, the difference $e$ can then be restored with $e\gets t_1/a$. When $\mathcal{P}_2$ receives $e$ from $\mathcal{P}_1$, the share of result $u_2$ can be directly obtained by computing $u_2\gets e+b$ and another difference $d$ can be restored with $d\gets t_2/u_2$. Lastly, $\mathcal{P}_1$ compute $u_1\gets d+a$ after he received $d$ from $\mathcal{P}_2$. Algorithm \ref{alg:secaddres} shows full details of this protocol and combines some computations.

It is worth mentioning that Algorithm \ref{alg:secaddres} includes two division operations and thus may cause the problem of division by zero. However, in fact, the cases that $a$ happens to be $0$ or $b$ happens to be $(c_1-x_1)/a$ are negligible. Another case needs to point out is that when the input value $x=0$ (i.e., $x_1=-x_2$), the share of result

\begin{align*}
    u_1 = & d+a = \frac{x_2-c_2}{u_2} + a \\
        = & \frac{-x_1-c_2}{\frac{x_1-c_1}{a}+b}+a = \frac{-ac+a^2b}{x_1-c_1+ab} = 0,
\end{align*}

\noindent is identically equal to zero. This means that once $\mathcal{P}_1$ get $u_1=0$, he can know that the secret value $x=0$, but it is the necessary price of performing this protocol. Even in the ideal world, one of the parties $\mathcal{P}_i$ will get knowledge of the secret value when it is zero. As illustrated in Figure \ref{fig:idealfunAR}, the ideal functionality $\mathcal{F}_{AR}$ which the additive resharing protocol wants to securely realize describes the behavior in the ideal world.

\begin{figure}
	\centering
	\fbox{%
		\parbox{0.95\linewidth}{%
			\begin{small}
				$\mathcal{F}_{AR}$ interacts with parties $\mathcal{P}_1$, $\mathcal{P}_2$, and adversary $\mathcal{S}$.
                \medskip
                \renewcommand{\theenumi}{\arabic{enumi}}
                \renewcommand{\labelenumi}{\theenumi.}
				\begin{enumerate}
                    \item $\mathcal{F}_{AR}$ receives $x_i$ from $\mathcal{P}_i$ for all $i=1,2$.
                    \item $\mathcal{F}_{AR}$ computes $x\gets x_1+x_2$.
                    \item $\mathcal{F}_{AR}$ chooses random nonzero value $u_2\in \mathbb{F}$.
                    \item $\mathcal{F}_{AR}$ computes $u_1\gets x/u_2$.
                    \item $\mathcal{F}_{AR}$ sends $u_1$ to $\mathcal{P}_1$ and $u_2$ to $\mathcal{P}_2$.
					\medskip	
                \end{enumerate}
			\end{small}
		}
	}
	\caption{Additive resharing ideal functionality $\mathcal{F}_{AR}$.}
	\label{fig:idealfunAR}
\end{figure}

\begin{theorem}
    Algorithm \ref{alg:secaddres} is correct and UC-secure in the honest-but-curios model.
\end{theorem}

\begin{IEEEproof}
    For correctness, there are
    \begin{align*}
        u = & u_1 \cdot u_2 = (d+a) \cdot (e+b) \\
          = & x_2-c_2 + ae + ab = x_2-c_2+x_1-c_1+ab \\
          = & x_1 + x_2 = x
    \end{align*}
    To prove security, we need to show that both views of $\mathcal{P}_1$ and $\mathcal{P}_2$ are simulatable. For $\mathcal{P}_2$, his incoming view is $(e)$, where $e=(x_1-c_1)/a$. And for $\mathcal{P}_1$, his incoming view is $(d)$, where $d=(x_2-c_2)/(e+b)$. Since the values $a,b,c_1,c_2$ are uniformly distributed and independent of any private input, due to Lemma \ref{lemma:add} and Lemma \ref{lemma:mul}, the incoming views of $\mathcal{P}_1$ and $\mathcal{P}_2$ are simulatable. Besides, it is trivial to see that the outputs of $\mathcal{P}_1$ and $\mathcal{P}_2$ can be also simulated.
\end{IEEEproof}

\begin{algorithm}[t]
    \caption{Secure Additive Resharing Protocol $\mss{u}\gets SecAddRes(\ass{x})$}\label{alg:secaddres}
    \begin{algorithmic}[1]
      \Require
        \Statex Shared secret $\ass{x}$ over ASS.
        \Statex Beaver triple: $a,b,\ass{c}$, where $c = a \cdot b$. The value $c$ is shared by ASS, meanwhile $a$ is held by $\mathcal{P}_1$ and $b$ is held by $\mathcal{P}_2$.
      \Ensure
        Shared secret $\mss{u}$ over MSS such that $u=x$.
      \State $\mathcal{P}_1$ computes $e=(x_1-c_1)/a$.
      \State $\mathcal{P}_1$ sends $e$ to $\mathcal{P}_2$.
      \State $\mathcal{P}_2$ computes $u_2\gets e + b, d\gets (x_2-c_2)/u_2$.
      \State $\mathcal{P}_2$ sends $d$ to $\mathcal{P}_1$.
      \State $\mathcal{P}_1$ computes $u_1\gets d+a$.
      \State \textbf{Return} $\mss{u}$
    \end{algorithmic}
\end{algorithm}

\subsubsection{Comparison}

It is easy to see that MSS allows share to reserve the sign of secret, thus secure comparison can be accomplished via additive resharing. As presented in Algorithm \ref{alg:seccom}, $\mathcal{P}_1$ and $\mathcal{P}_2$ compute the difference $\ass{d}=\ass{x}-\ass{y}$ at first, then call secure additive resharing protocol to get the difference $\mss{u}$ over MSS, and after that, the sign of $\mss{u}$ will be revealed between each party. Since there exists round-off error in practice, two floating-point numbers are almost impossible to be completely equal. A precision threshold $\Delta$ could be set to determine whether two floats are equal, i.e., whether their difference is reduced to zero. When the signs of all multiplicative shares are public, parties will arrive at a consensus over the comparison of two numbers.

\begin{theorem}
    Algorithm \ref{alg:seccom} is correct and UC-secure in honest-but-curios model.
\end{theorem}

\begin{IEEEproof}
    It is trivial to prove the correctness and security of Algorithm \ref{alg:seccom}, since those of \texttt{SecAddRes} protocol has been proven.
\end{IEEEproof}

\begin{algorithm}[t]
    \caption{Secure Comparison Protocol $SecCom(\ass{x},\ass{y})$}\label{alg:seccom}
    \begin{algorithmic}[1]
      \Require
        Shared secrets $\ass{x}$ and $\ass{y}$.
      \Ensure
        $0$ if $x==y$, $1$ if $x>y$, $-1$ if $x<y$.
      \State $\mathcal{P}_1$ computes $d_1\gets x_1-y_1$.
      \State $\mathcal{P}_2$ computes $d_2\gets x_2-y_2$.
      \State $\mathcal{P}_1$ and $\mathcal{P}_2$ collaboratively compute $\mss{u}\gets SecAddRes(\ass{d})$.
      \State $\mathcal{P}_1$ and $\mathcal{P}_2$ reveal the signs of $u_1$ and $u_2$.
      \If{$u_1==0$ or $u_2==0$}
      \State \textbf{Return} $0$.
      \ElsIf{($u_1>0$ and $u_2>0$) or ($u_1<0$ and $u_2<0$)}
      \State \textbf{Return} $1$.
      \Else
      \State \textbf{Return} $-1$.
      \EndIf
    \end{algorithmic}
\end{algorithm}

\subsection{Nonlinear Functions}

Switching shared secret between two sharing forms as well as some identities makes possible secure protocols for exponentiation, logarithm, and power without any approximate or bit-level operation. Besides, the division can be seen as a special case of power.

\subsubsection{Exponentiation}

Secure exponentiation is to compute $\ass{f}\gets a^{\ass{x}}$, where $a$ is a public base number and is usually specified as $e$. According to the identity

\begin{equation}
    a^{x_1+x_2}=a^{x_1}\cdot a^{x_2},
\end{equation}

\noindent when parties perform exponential operation on their shares, the problem can be transformed into the problem of converting multiplicative shares which has been solved by \texttt{SecMulRes}. Without loss of generality, Algorithm \ref{alg:secexp} takes a secret exponent $\ass{x}$ and a public base number $a$ as input, and returns the shared result $\ass{a^x}$.

In addition, when base number $a$ is negative, an inappropriate exponent such as $\sfrac{1}{2}$ will lead to a mistake in computing. However, such case should be avoided in designing function, not be guaranteed of this algorithm.

\begin{theorem}
    Algorithm \ref{alg:secexp} is correct and UC-secure in honest-but-curios model.
\end{theorem}

\begin{IEEEproof}
    According to correctness of \texttt{SecMulRes}, we have
    \begin{equation*}
        f_1+f_2=u_1\cdot u_2,
    \end{equation*}
    \noindent therefore, the correctness of this algorithm can be proven as follow
    \begin{equation*}
        f=f_1+f_2=u_1\cdot u_2=a^{x_1}\cdot a^{x_2}=a^{x_1+x_2}=a^x.
    \end{equation*}
    Since \texttt{SecMulRes} is secure and this algorithm contains no additional communication, security is trivial.
\end{IEEEproof}

\begin{algorithm}[t]
    \caption{Secure Exponentiation Protocol $\ass{f}\gets SecExp(\ass{x},a)$}\label{alg:secexp}
    \begin{algorithmic}[1]
      \Require
        Shared secret $\ass{x}$ and public base number $a$.
      \Ensure
        Shared secret $\ass{f}$ such that $f=a^x$.
      \State $\mathcal{P}_1$ computes $u_1\gets a^{x_1}$.
      \State $\mathcal{P}_2$ computes $u_2\gets a^{x_2}$.
      \State $\mathcal{P}_1$ and $\mathcal{P}_2$ collaboratively compute $\ass{f}\gets SecMulRes(\mss{u})$.
      \State \textbf{Return} $\ass{f}$.
    \end{algorithmic}
\end{algorithm}

\subsubsection{Logarithm}

As the inverse of exponentiation, secure logarithm is to compute $\ass{f}\gets \log_a\ass{x}$, where $a$ is a base number. Similar to exponentiation, logarithm has its own ''product-to-sum'' identity

\begin{equation}
    \log_a (u_1\cdot u_2)=\log_a u_1 + \log_a u_2.
\end{equation}

From this point, parties first call the protocol \texttt{SecAddRes} to convert additive shares into multiplicative shares and then perform logarithmic operation on each share. Since logarithmic function requires positive input, we have to make the share absolute in advance. It means that what we actually designed is the secure function $\ass{f}\gets\log_a|\ass{x}|$, and which is more reasonable. Note that the steps of \texttt{SecLog} are exactly the opposite of \texttt{SecExp}, in which the steps of \texttt{SecAddRes} and \texttt{SecMulRes} they called are also opposite.

\begin{theorem}
    Algorithm \ref{alg:seclog} is correct and UC-secure in honest-but-curios model.
\end{theorem}

\begin{IEEEproof}
    For correctness, there are
    \begin{equation*}
        f=f_1+f_2=\log_a|u_1|+\log_a|u_2|=\log_a|u_1\cdot u_2|,
    \end{equation*}
    \noindent due to the correctness of \texttt{SecAddRes}, we have
    \begin{equation*}
        f=\log_a|u_1\cdot u_2|=\log_a|x_1+x_2|=\log_a|x|.
    \end{equation*}
    Since \texttt{SecAddRes} is secure and this algorithm contains no additional communication, security is trivial.
\end{IEEEproof}

\begin{algorithm}[t]
    \caption{Secure Logarithm Protocol $\ass{f}\gets SecLog(\ass{x},a)$}\label{alg:seclog}
    \begin{algorithmic}[1]
      \Require
        Shared secret $\ass{x}$ and public base number $a$.
      \Ensure
        Shared secret $\ass{f}$ such that $f=\log_a|x|$.
      \State $\mathcal{P}_1$ and $\mathcal{P}_2$ collaboratively compute $\mss{u}\gets SecAddRes(\ass{x})$.
      \State $\mathcal{P}_1$ computes $f_1\gets \log_a|u_1|$.
      \State $\mathcal{P}_2$ computes $f_2\gets \log_a|u_2|$.
      \State \textbf{Return} $\ass{f}$.
    \end{algorithmic}
\end{algorithm}

\subsubsection{Power}

Secure power is to compute $\ass{f}\gets \ass{x}^\alpha$, where $\alpha$ is a public integer. Inspired by the identity

\begin{equation}
    \label{eq:pow}
    (u_1\cdot u_2)^\alpha=u_1^\alpha\cdot u_2^\alpha,
\end{equation}

\noindent we can first turn shares into multiplicative shares, then perform power operation on each share, and finally, turn shares back to additive. In fact, the above identity also holds for multiple multipliers, in other words, multiplication and power are to MSS what linear is to ASS. For $n$ multipliers, we have

\begin{equation}
    (u_1^1u_2^1u_1^2u_2^2 \cdots u_1^nu_2^n)^\alpha=(u_1^1u_1^2\cdots u_1^n)^\alpha\cdot (u_2^1u_2^2\cdots u_2^n)^\alpha.
\end{equation}

Therefore, as shown in Algorithm \ref{alg:secpow}, we design secure power protocol which implements secure function $\ass{f}\gets \prod_{j=1}^n{\ass{x}^j}^{\alpha^j}$. Note that division is the special case of power, i.e., \textsf{SecPow($\ass{x}$,$\ass{y}$,$1$,$-1$)}, and product is also the special case, i.e., \textsf{SecPow($\ass{x^1}$,$\ass{x^2}$,$\cdots$,$\ass{x^n}$,$1$,$1$,$\cdots$,$1$)}.

In addition, when the exponent $\alpha$ is a real number or even a shared secret, this algorithm no longer applies. We discussed both cases in the Appendix.

\begin{theorem}
    Algorithm \ref{alg:secpow} is correct and UC-secure in honest-but-curios model.
\end{theorem}

\begin{IEEEproof}
    For correctness, there are
    \begin{equation*}
        f=f_1+f_2=v_1\cdot v_2=\prod_{j=1}^n {u_1^j}^{\alpha^j}\cdot {u_2^j}^{\alpha^j}=\prod_{j=1}^n {u^j}^{\alpha^j}=\prod_{j=1}^n {x^j}^{\alpha^j}.
    \end{equation*}
    Since \texttt{SecAddRes} and \texttt{SecMulRes} are secure and this algorithm contains no additional communication, security is trivial.
\end{IEEEproof}

\begin{algorithm}[t]
    \caption{Secure Power Protocol $\ass{f}\gets SecPow(\ass{x^1},\ass{x^2},\cdots,\ass{x^n},\alpha^1,\alpha^2,\cdots,\alpha^n)$}\label{alg:secpow}
    \begin{algorithmic}[1]
      \Require
        Shared secrets $\ass{x^j}$ and public integers $\alpha^j$ for all $j\in \{1,2,\cdots,n\}$.
      \Ensure
        Shared secret $\ass{f}$ such that $f=\prod_{j=1}^n {x^j}^{\alpha^j}$.
      \State $\mathcal{P}_1$ and $\mathcal{P}_2$ collaboratively compute $\mss{u^j}\gets SecAddRes(\ass{x^j})$ for all $j\in \{1,2,\cdots,n\}$ in parallel.
      \State $\mathcal{P}_1$ computes $v_1\gets \prod_{j=1}^n {u_1^j}^{\alpha^j}$.
      \State $\mathcal{P}_2$ computes $v_2\gets \prod_{j=1}^n {u_2^j}^{\alpha^j}$.
      \State $\mathcal{P}_1$ and $\mathcal{P}_2$ collaboratively compute $\ass{f}\gets SecMulRes(\mss{v})$.
      \State \textbf{Return} $\ass{f}$.
    \end{algorithmic}
\end{algorithm}

\subsubsection{Trigonometric Functions}

Trigonometric functions which contain sine, cosine, tangent, cotangent, cosecant, and secant describe the relationship of an angle and two edges of a right triangle. We first design secure protocols for two basic functions: sine and cosine, then discuss the remains.

Secure sine is to compute $\ass{f}\gets \sin\ass{x}$. Considering the angle sum and difference identity

\begin{equation}
    \sin(x_1+x_2)=\sin x_1 \cdot \cos x_2 + \cos x_1 \cdot \sin x_2,
\end{equation}

\noindent the problem is transformed into how to convert shares. Algorithm \ref{alg:secsin} first computes the sine and cosine of each share, then divides above four trigonometric results into two groups of multiplicative shares, and the rest is to call \texttt{SecMulRes} for each group in parallel.

Secure cosine is to compute $\ass{f}\gets \cos\ass{x}$. Similar to Algorithm \ref{alg:secsin}, it is easy to design secure cosine protocol that securely implements cosine according to the identity

\begin{equation}
    \cos(x_1+x_2)=\cos x_1 \cdot \cos x_2 - \sin x_1 \cdot \sin x_2.
\end{equation}

\noindent Algorithm \ref{alg:seccos} shows the details.

\begin{theorem}
    Algorithm \ref{alg:secsin} and Algorithm \ref{alg:seccos} are correct and secure in honest-but-curios model.
\end{theorem}

\begin{IEEEproof}
    For the correctness of Algorithm \ref{alg:secsin}, there are
    \begin{align*}
        f = & f_1 + f_2 = f_1^m + f_1^n + f_2^m + f_2^n = f^m + f^n \\
          = & m + n = m_1m_2 + n_1n_2 = \sin x_1 \cdot \cos x_2 + \cos x_1 \cdot \sin x_2 \\
          = & \sin(x_1+x_2) = \sin x.
    \end{align*}
    \noindent Referring to the above proof, it is trivial to prove the correctness of Algorithm \ref{alg:seccos}.
    Since \texttt{SecMulRes} is secure and Algorithm \ref{alg:secsin} and Algorithm \ref{alg:seccos} contain no additional communication, security of both is trivial.
\end{IEEEproof}

\begin{algorithm}[t]
    \caption{Secure Sine Protocol $\ass{f}\gets SecSin(\ass{x})$}\label{alg:secsin}
    \begin{algorithmic}[1]
      \Require
        Shared secret $\ass{x}$.
      \Ensure
        Shared secret $\ass{f}$ such that $f=\sin x$.
      \State $\mathcal{P}_1$ computes $m_1\gets\sin x_1, n_1\gets \cos x_1$.
      \State $\mathcal{P}_2$ computes $n_2\gets\sin x_2, m_2\gets \cos x_2$.
      \State $\mathcal{P}_1$ and $\mathcal{P}_2$ collaboratively compute $\ass{f^m}\gets SecMulRes(\mss{m}), \ass{f^n}\gets SecMulRes(\mss{n})$ in parallel.
      \State $\mathcal{P}_1$ computes $f_1\gets f_1^m+f_1^n$.
      \State $\mathcal{P}_2$ computes $f_2\gets f_2^m+f_2^n$.
      \State \textbf{Return} $\ass{f}$.
    \end{algorithmic}
\end{algorithm}

\begin{algorithm}[t]
    \caption{Secure Cosine Protocol $\ass{f}\gets SecCos(\ass{x})$}\label{alg:seccos}
    \begin{algorithmic}[1]
      \Require
        Shared secret $\ass{x}$.
      \Ensure
        Shared secret $\ass{f}$ such that $f=\cos x$.
      \State $\mathcal{P}_1$ computes $m_1\gets\sin x_1, n_1\gets \cos x_1$.
      \State $\mathcal{P}_2$ computes $m_2\gets\sin x_2, n_2\gets \cos x_2$.
      \State $\mathcal{P}_1$ and $\mathcal{P}_2$ collaboratively compute $\ass{f^m}\gets SecMulRes(\mss{m}), \ass{f^n}\gets SecMulRes(\mss{n})$ in parallel.
      \State $\mathcal{P}_1$ computes $f_1\gets f_1^n-f_1^m$.
      \State $\mathcal{P}_2$ computes $f_2\gets f_2^n-f_2^m$.
      \State \textbf{Return} $\ass{f}$.
    \end{algorithmic}
\end{algorithm}

Tangent, cotangent, cosecant, and secant could be defined by sine and cosine as $\sfrac{sine}{cosine}$, $\sfrac{cosine}{sine}$, $\sfrac{1}{sine}$, and $\sfrac{1}{cosine}$ respectively. Therefore, constructing secure protocols for these functions is merely a combination of \texttt{SecSin}, \texttt{SecCos}, and \texttt{SecPow}, and it is trivial to prove the correctness and security of them.

\subsection{Inverse Trigonometric Functions}

The inverse trigonometric functions is the final problem to integrity basic elementary functions. However, there is no angle sum and difference identity for inverse trigonometric functions. Therefore, we have to settle for second best by using Taylor series. For instance, $arcsin(x)$ can be calculated using Maclaurin series:

\begin{equation}
  arcsin(x) = x + \left(\frac{1}{2}\right)\frac{x^3}{3} + \left(\frac{1\cdot 3}{2\cdot 4}\right)\frac{x^5}{5} + \left(\frac{1\cdot 3\cdot 5}{2\cdot 4 \cdot 6}\right)\frac{x^7}{7} + \cdots
\end{equation}

\noindent Since secure power protocol is with constant round complexities, high order Taylor series (Maclaurin series) will not increase additional interaction between parties. As a result, we could increase the order of Taylor series as much as possible to improve accuracy.

\section{Complexity}\label{sec:complexity}

Table \ref{tab:complexity} summarized round complexities and communications of all the protocols in this paper. We can see that all proposed protocols have low constant round complexities and low communications, which is suitable for practice.

The round complexities of \texttt{SecCom} are $3$ which contains $2$ rounds of \texttt{SecAddRes} and $1$ round of revealing signs. As the sign of secret over MSS is precisely in ASS form, parties do not need to know the exact result of comparison in some scenarios. Consequently, \texttt{SecAddRes} can be regarded as secure comparison protocol with semaphore, and its round complexities are $2$.

Secure product is available in two methods: 1) recursively splitting input sequences in half and then computing each pair in parallel; 2) as a special case of power. The Round complexities of the two methods are $\lceil \log_2 n \rceil$ and the constant $3$, and communications are $(4n-4)l$ and $(2n+2)l$, respectively. Hence the round complexities of product are $min(3,\lceil \log_2 n \rceil)$ by selecting the appropriate method according to the number of inputs.

The round complexities of secure protocols for tangent, cotangent, cosecant and secant defined by sine and cosine are $4$, while what using the angle sum and difference identities are $4$ as well.

\begin{table}
    \centering
    \caption{Complexities of protocols \protect \\ (Here, $l$ is the bit-width of the element in practice, and  $n$ denotes the number of inputs)}
	\label{tab:complexity}
    \begin{tabular}{@{}lcc@{}}
    \toprule
    Protocol                         & Round Complexities & Communications      \\ \midrule
    \texttt{SecMul}                  & $1$                & $4l$                \\
    \texttt{SecMulRes}               & $1$                & $2l$                \\
    \texttt{SecAddRes}               & $2$                & $2l$                \\
    \texttt{SecCom}                  & $3$                & $2l+2$              \\
    \texttt{SecExp}                  & $1$                & $2l$                \\
    \texttt{SecLog}                  & $2$                & $2l$                \\
    \texttt{SecPow}                  & $3$                & $(2n+2)l$           \\
    \texttt{SecSin}, \texttt{SecCos} & $1$                & $4l$                \\
    Division                         & $3$                & $6l$                \\
    Product                          & $min(3,\lceil \log_2 n \rceil)$   & $(2n+2)l$, $(4n-4)l$ \\
    Tagent, Cotangent                & $4$                & $14l$               \\
    Secant, Cosecant                 & $4$                & $12l$               \\ \bottomrule
    \end{tabular}
\end{table}

Table \ref{tab:com} compared the complexities of secure protocols for comparison and division with previous work, and showed that our scheme is more efficient.

\begin{table*}[ht]
    \centering
    \caption{Comparison of Round Complexities and Communications \protect \\ (Here, $l$ is the bit-width, $l'=\log_2 l$, and the values of $m, n'$ determine the accuracy of division)}
    \label{tab:com}
    \begin{tabular}{|c|c|c|c|c|}
    \hline
    \multirow{2}{*}{} & \multicolumn{2}{c|}{Comparison} & \multicolumn{2}{c|}{Division} \\ \cline{2-5} 
    Scheme            & Rounds   & Comm.       & Rounds     & Comm.   \\ \hline
    \cite{nishide2007multiparty}            & $15$     & $279l+5$             & -          & -                \\ \hline
    \cite{bogdanov2012high}            & $l'+3$   & $5{l'}^2+12(l'+1)l$  & $4l'+9$    & $2ml+6ml'+39l'l+35l'n'+126l+32n'+24$                \\ \hline
    \cite{huang2019lightweight}            & $l+3$    & $10l-2$              & -          & -                \\ \hline
    Ours              & $3$      & $2l+2$               & $3$        & $6l$             \\ \hline
    \end{tabular}
\end{table*}

\section{Conclusions}\label{sec:conclusions}

In this paper, we firstly proposed two basic protocols, multiplicative resharing and additive resharing, which can switch shared secrets. And then, based on this ability, protocols for comparison, exponentiation, logarithm, power, product, division, and trigonometric functions were proposed, where product and division derived from power. Additionally, we discussed the cases about power with real exponent or with secret exponent in Appendix, and constructed protocols for them. All the protocols proposed in this paper are composed of arithmetic operations and contain no bit-level operation with low constant round complexities, and have been proven secure in honest-but-curios model.

Specially, secure power protocol with constant round complexities provides the possibility of higher order Taylor series which can approximate most functions that include those not implemented in this paper and will benefit from higher order. Meanwhile, secure sine and cosine protocols provide the possibility of efficient Fourier series which can approximate an arbitrary function in a period.

\section*{Acknowledgements}

This work is supported in part by the National Natural Science Foundation of China under grant numbers 61702276, 61672294, 61502242, in part by the Priority Academic Program Development of Jiangsu Higher Education Institutions (PAPD) fund.

\bibliographystyle{IEEEtran}

\bibliography{IEEEabrv, ref}


\begin{IEEEbiography}[{\includegraphics[width=1in,height=1.25in,clip,keepaspectratio]{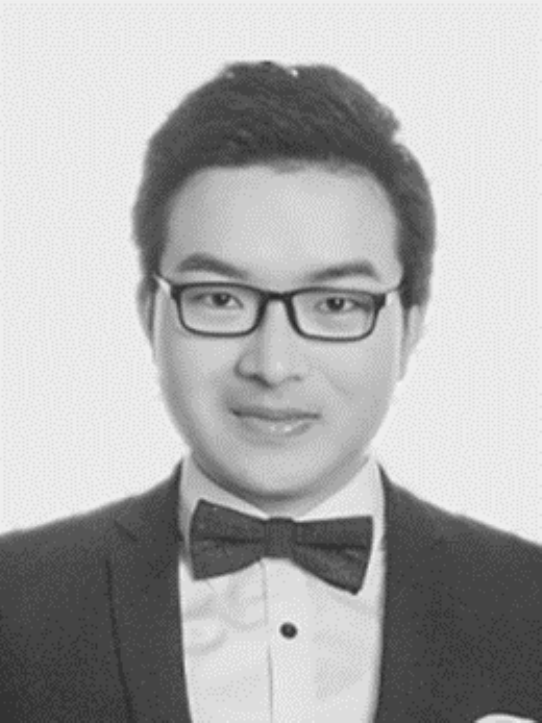}}]{Lizhi Xiong}
  received Ph.D. degree in Communication and Information System from Wuhan University, China in 2016. From 2014 to 2015, he was a Joint-Ph.D. student with Electrical and Computer Engineering, New Jersey University of Technology, New Jersey, USA. He is currently an Associate Professor with School of Computer and Software, Nanjing University of Information Science and Technology, Nanjing, China. His main research interests include privacy-preserving computation, information hiding, and multimedia security.
\end{IEEEbiography}

\begin{IEEEbiography}[{\includegraphics[width=1in,height=1.25in,clip, keepaspectratio]{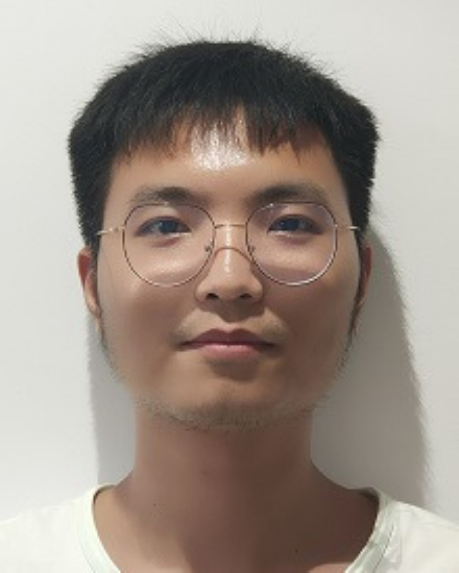}}]{Wenhao Zhou}
  is currently pursuing his M.S. degree in the School of Computer and Software, Nanjing University of Information Science and Technology, China. His research interests include secure multiparty computation and privacy-preserving computation.
\end{IEEEbiography}

\begin{IEEEbiography}[{\includegraphics[width=1in,height=1.25in,clip,keepaspectratio]{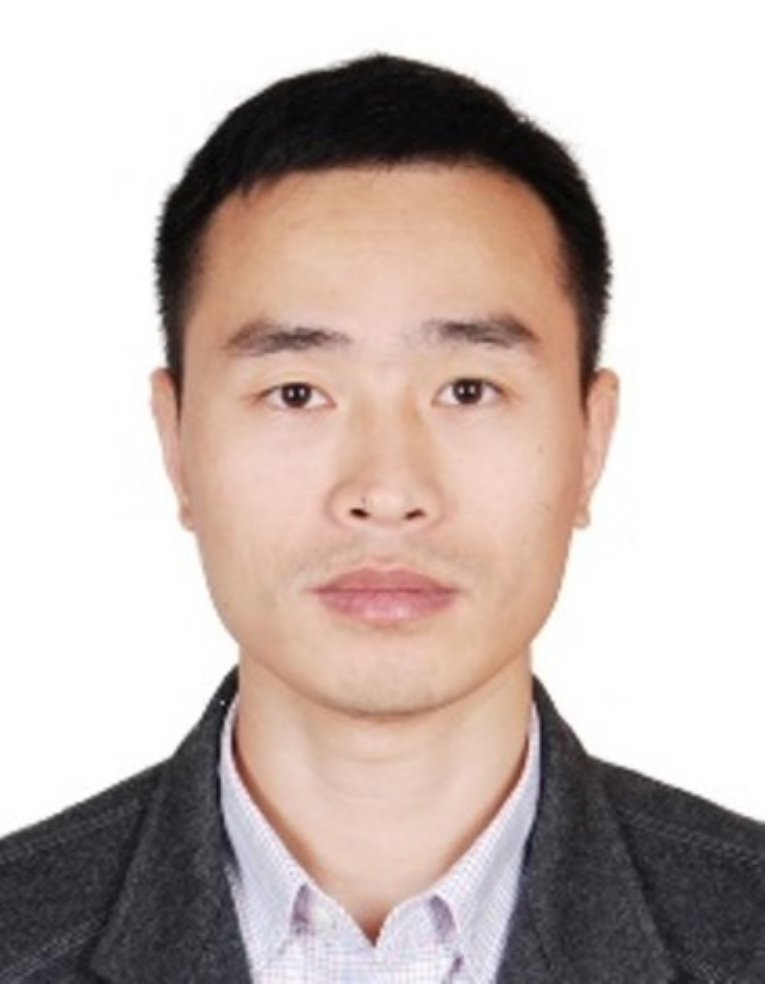}}]{Zhihua Xia}
  received his B.S. degree in Hunan City University, China, in 2006, and the Ph.D. degree in computer science and technology from Hunan University, China, in 2011.He is currently an associate professor with the School of Computer and Software, Nanjing University of Information Science and Technology, China. He was a visiting professor with the Sungkyunkwan University, Korea, 2016. His research interests include cloud computing security and digital forensic. He is a member of the IEEE.
\end{IEEEbiography}

\begin{IEEEbiography}[{\includegraphics[width=1in,height=1.25in,clip, keepaspectratio]{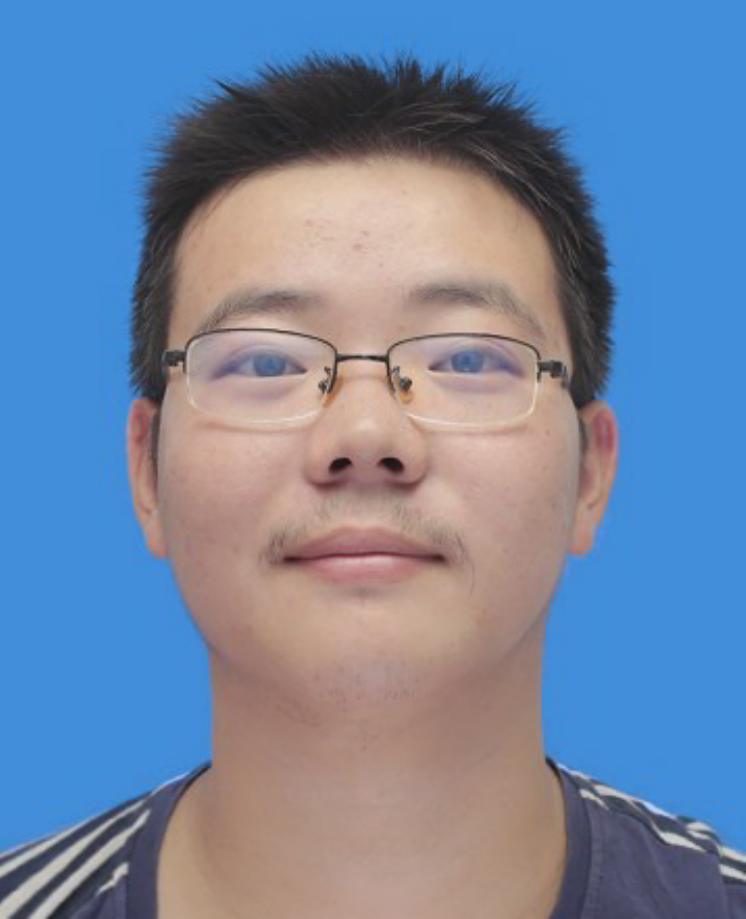}}]{Qi Gu}
  is currently pursuing his M.S. degree in the School of Computer and Software, Nanjing University of Information Science and Technology, China. His research interests include functional encryption, image retrieval and nearest neighbor search.
\end{IEEEbiography}

\begin{IEEEbiography}[{\includegraphics[width=1in,height=1.25in,clip,keepaspectratio]{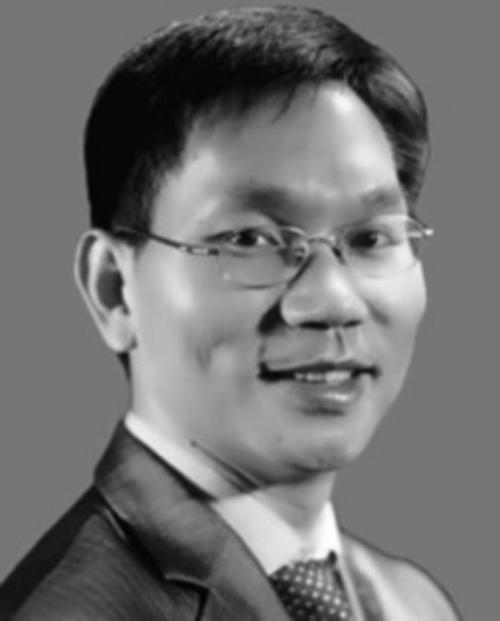}}]{Jian Weng}
  received the B.S. and M.S. degrees in computer science and engineering from South China University of Technology, Guangzhou, China, in 2000 and 2004, respectively, and the Ph.D. degree in computer science and engineering from Shanghai Jiao Tong University, Shanghai, China, in 2008. From 2008 to 2010, he held a Postdoctoral position with the School of Information Systems, Singapore Management University. He is currently a Professor and the Dean with the College of Information Science and Technology, Jinan University, Guangzhou,China. He has authored or coauthored more than 100 papers in cryptography and security conferences and journals, such as CRYPTO, EUROCRYPT, ASIACRYPT, TCC, PKC, TPAMI, TIFS, and TDSC. His research interests include public key cryptography, cloud security, and blockchain. He was the PC Co-Chairs or PC Member for more than 30 international conferences. He also serves as an Associate Editor for the IEEE TRANSACTIONS ON VEHICULART ECHNOLOGY. 
\end{IEEEbiography}

\newpage

\appendix[Power with Real or Secret Exponent]

When the exponent is real as well as the base is negative, Algorithm \ref{alg:secpow} will produce wrong result, since the identity \ref{eq:pow} doesn't hold under this case. However, we note that the identity

\begin{equation}
    u^\alpha = {(-1)}^\alpha {|u|}^\alpha,
\end{equation}

\noindent where $u$ is a negative number. This provides us with an idea: firstly, compute ${|\ass{x}|}^\alpha$ as with Algorithm \ref{alg:secpow}, secondly, multiply each share by ${(-1)}^\alpha$ if $x$ is negative. We thus constructed secure power with real exponent protocol as described in Algorithm \ref{alg:secpre}, where to simplify the problem, the case with one multiplier is considered. The protocol is with complexity of $4$ rounds and $4l+2$ communications.

Besides, it should be emphasized that the coefficient ${(-1)}^\alpha$ usually not only directs the real field to the complex field, but also leads to multiple results. Although all protocols in this paper also available in the complex field with necessary adjustments, many practical applications do not need to enter the complex field. In this case, it only needs to compute ${|\ass{x}|}^\alpha$ without multiplying ${(-1)}^\alpha$. For example, batch normalization \cite{ioffe2015batch} involving root operation assumes that the input is positive.

\begin{theorem}
    Algorithm \ref{alg:secpre} is correct and UC-secure in honest-but-curios model.
\end{theorem}

\begin{IEEEproof}
    For correctness, if $x$ is negative, there are
    \begin{equation*}
        f={(-1)}^\alpha (f_1+f_2)={(-1)}^\alpha(v_1v_2)={(-1)}^\alpha{|u|}^\alpha=x^\alpha.
    \end{equation*}
    \noindent If not, it is same as Algorithm \ref{alg:secpow}.
    Since \texttt{SecAddRes}, \texttt{SecCom}, \texttt{SecMulRes} are secure and this algorithm contains no additional communication, security is trivial.
\end{IEEEproof}

\begin{algorithm}
    \caption{Secure Power with Real Exponent Protocol $\ass{f}\gets SecPRE(\ass{x},\alpha)$}\label{alg:secpre}
    \begin{algorithmic}[1]
      \Require
        Shared secret $\ass{x}$ and public real $\alpha$.
      \Ensure
        Shared secret $\ass{f}$ such that $f=x^\alpha$.
      \State $\mathcal{P}_1$ and $\mathcal{P}_2$ collaboratively compute $\mss{u}\gets SecAddRes(\ass{x})$.
      \State $\mathcal{P}_1$ computes $v_1\gets {|u_1|}^\alpha$.
      \State $\mathcal{P}_2$ computes $v_2\gets {|u_2|}^\alpha$.
      \State $\mathcal{P}_1$ and $\mathcal{P}_2$ collaboratively compute $\ass{f}\gets SecMulRes(\mss{v})$.
      \If{$SecCom(\ass{x},\ass{0})==-1$}
      \State $\mathcal{P}_1$ computes $f_1\gets f_1\cdot {(-1)}^\alpha$.
      \State $\mathcal{P}_2$ computes $f_2\gets f_2\cdot {(-1)}^\alpha$.
      \EndIf
      \State \textbf{Return} $\ass{f}$.
    \end{algorithmic}
\end{algorithm}

We now discuss the case that even the exponent is a shared secret. Since share cannot be exponentiated directly, we need to use the identity

\begin{equation}
    \alpha \ln u = \ln u^\alpha,
\end{equation}

\noindent to convert exponentiation into multiplication. In addition, as a shared secret, the exponent is almost impossible to be an integer, thereby we should also classify the positive and negative of the base. Algorithm \ref{alg:secpse} shows the details of secure power with secret exponent protocol. The protocol requires complexity of $5$ rounds and $8l+2$ communications.

\begin{theorem}
    Algorithm \ref{alg:secpse} is correct and UC-secure in honest-but-curios model.
\end{theorem}

\begin{IEEEproof}
    For correctness, if $x$ is negative, there are
    \begin{align*}
        f = & f_1+f_2=v_1\cdot v_2=e^{s_1}\cdot (-1)^{y_1} \cdot e^{s_2} \cdot (-1)^{y_2} = (-1)^ye^s \\
          = & (-1)^ye^{ty} = (-1)^ye^{y\ln |x|} = (-1)^y{|x|}^y = x^y.
    \end{align*}
    \noindent Otherwise, it is trivial.
    Since this algorithm consists of only secure protocols and contains no additional communication, security is trivial.
\end{IEEEproof}

\begin{algorithm}
    \caption{Secure Power with Secret Exponent Protocol $\ass{f}\gets SecPSE(\ass{x},\ass{y})$}\label{alg:secpse}
    \begin{algorithmic}[1]
      \Require
        Shared secrets $\ass{x}$ and $\ass{y}$.
      \Ensure
        Shared secret $\ass{f}$ such that $f=x^y$.
      \State $\mathcal{P}_1$ and $\mathcal{P}_2$ collaboratively compute $\ass{t}\gets SecLog(\ass{x})$.
      \State $\mathcal{P}_1$ and $\mathcal{P}_2$ collaboratively compute $\ass{s}\gets SecMul(\ass{t},\ass{y})$.
      \If{$SecCom(\ass{x},\ass{0})==1$}
      \State $\mathcal{P}_1$ and $\mathcal{P}_2$ collaboratively compute $\ass{f}\gets SecExp(\ass{s})$.
      \State \textbf{Return} $\ass{f}$.
      \ElsIf{$SecCom(\ass{x},\ass{0})==-1$}
      \State $\mathcal{P}_1$ computes $v_1\gets e^{s_1}\cdot (-1)^{y_1}$.
      \State $\mathcal{P}_2$ computes $v_2\gets e^{s_2}\cdot (-1)^{y_2}$.
      \State $\mathcal{P}_1$ and $\mathcal{P}_2$ collaboratively compute $\ass{f}\gets SecMulRes(\mss{v})$.
      \State \textbf{Return} $\ass{f}$.
      \Else
      \State \textbf{Return} $\ass{0}$.
      \EndIf
    \end{algorithmic}
\end{algorithm}

\end{document}